\documentclass{eas}
\usepackage{graphicx}
\begin{document}

%%-----------------------------
%%      the top matter
%%-----------------------------
\title{Atmospheric parameter determination for massive stars via
  non-LTE spectrum analysis} 
\author{Mar\'ia-Fernanda Nieva}\address{MPI for Astrophysics, 
Karl-Schwarzschild-Str. 1, D-85741 Garching, Germany}
\author{Norbert Przybilla}\address{Dr. Remeis-Sternwarte Bamberg \&
ECAP, Astronomisches Institut der Universit\"at Erlangen-N\"urn\-berg, 
Sternwartstrasse 7, D-96049 Bamberg, Germany}

\runningtitle{Nieva \& Przybilla: Atmospheric Parameters from Non-LTE Analysis}

\begin{abstract}
We describe a self-consistent spectrum analysis technique employing non-LTE
line formation, which allows precise atmospheric parameters of massive stars 
to be derived: 1$\sigma$-uncer\-tainties as low as $\sim$1\% in 
effective temperature and $\sim$0.05--0.10\,dex in surface gravity can be
achieved. 
Special emphasis is given to the minimisation of the
main sources of systematic errors in the atmospheric model computation, 
the observed spectra and the quantitative spectral analysis.
Examples of applications are discussed for OB-type stars near the main sequence and their
evolved progeny, the BA-type supergiants, covering masses of $\sim$8 to 25\,M$_\odot$ and 
a range in effective temperature from $\sim$8\,000 to 35\,000\,K.
Relaxing the assumption of local thermodynamic equilibrium 
in stellar spectral synthesis has been shown to be decisive for
improving the accuracy of quantitative analyses. Despite the present examples, which concentrate 
on hot, massive stars, the same philosophy can be applied to line-formation
calculations for all types of stars, including cooler objects like the Sun, once
the underlying stellar atmospheric physics is reproduced consistently.

\end{abstract}
\maketitle
%%-----------------------------
%%      your text
%%-----------------------------
\section{Introduction}
Many fields of contemporary astrophysics require a highly-precise determination
of basic stellar parameters. A prime example is the investigation of stellar
structure and evolution, where asteroseismology has opened up a window to
study the governing physical processes in detail (e.g.~Aerts et
al.~\cite{aerts08}). Other examples are
observational constraints on planet-formation theories by analyses of
planet host stars (e.g.~Neves et al.~\cite{neves09}). Predictions of 
super-/hyper--nova models may be tested
via analyses of secondary stars in low-mass X-ray
binaries (e.g.~Gonz\'alez Hern\'andez et al.~\cite{GRI08}) and
(hyper-)runaway stars (Przybilla et al.~\cite{PNHB08}).
Tight constraints on Galactochemical evolution theories may be derived 
using stars as tracers of the
variation of chemical composition with time and location in the Galactic disk
(e.g.~Fuhrmann~\cite{fuhrmann08}; Przybilla~\cite{P08}).

Precise stellar parameter determinations are decisive in describing the
stellar atmospheric structure correctly, i.e.~the temperature gradient and density
stratification. This is crucial for the regions where the continuum 
and spectral lines of interest are formed. A fundamental problem is that a
direct determination of stellar parameters is usually not possible. 
Quantitative spectral analysis has to rely on the comparison of synthetic spectra 
with observation, which in turn guides computations of improved models. In principle, 
this requires an iterative approach. In practice, however, approximate 
methods for the parameter determination are favoured on most occasions to shorten 
the process, by e.g.~use of photometric indicators or comparison with prescribed 
(and thus necessarily limited) grids of models. Unfortunately, the gain in efficiency is 
often accompanied by a loss in accuracy. 

Quantitative spectroscopy is highly prone to systematic errors.
The analysis methodology, the quality of the observed spectra and 
the details of the assumptions made in the modelling, when taken together, determine
the accuracy of a work. In particular, relaxing the assumption of local 
thermodynamic equilibrium (LTE) for the spectrum synthesis is 
crucial for improving the accuracy of stellar analyses.

In what follows we describe the basic requirements for a highly accurate
determination of stellar atmospheric parameters by utilising all available
lines of hydrogen and helium, plus multiple metal ionization equilibria, in non-LTE.
We concentrate on OB-type stars near the main sequence and on their evolved progeny,
BA-type supergiants. The stellar mass range between $\sim$8 and 25\,M$_\odot$
is covered for effective temperatures from $\sim$8000 to 35\,000\,K. 
The models and analysis technique have been developed and thoroughly tested during the past
years (Przybilla et al.~\cite{PBBK06}, PBBK06 henceforth; Nieva \&
Przybilla~\cite{NP07,NP08}, NP07/NP08; Przybilla,
Nieva \& Butler~\cite{PNB08}, PNB08).
One should keep in mind that the same philosophy can, in principle, be applied 
to any type of star once the underlying stellar atmospheric physics is reproduced 
consistently.

We start with an overview of 
requirements on atmospheric models, the spectrum synthesis and the
quality of observed spectra for a reliable quantitative analysis
(Sects.~\ref{modelrequirements} \& \ref{observations}).
Then, a self-consistent spectrum analysis technique is described
(Sect.~\ref{analysis}). Concrete examples of stellar parameter
determinations via non-LTE spectral analyses are discussed
(Sect.~\ref{examples}). Finally, common sources of systematic error are 
briefly addressed (Sect.~\ref{errors}).

\section{Model Requirements for the Spectral Analysis}\label{modelrequirements}
An implicit requirement for high-precision quantitative spectrum analysis is the
availability of model atmosphere and line-formation codes that account for
all physics relevant to the objects of interest. All
assumptions/approximations made to simplify the modelling need to be 
tested thoroughly to avoid fundamental sources of systematic
effects in the analysis. Familiarity with the codes and the physics
behind them are essential for successful work.

On the microscopic scale, the thermodynamic state of the plasma in a stellar 
atmosphere is determined by two competitive processes. Collisions act to
establish detailed equilibrium {\it locally}, while radiative processes are 
{\it non-local} in character\footnote{Photons can travel large distances in a
stellar atmosphere before interacting with the particles, i.e.~the radiation
field couples the plasma conditions from different depths of the atmosphere.}, thus driving the 
plasma out of detailed equilibrium. 
Local thermodynamic equilibrium is established only
when either collisions dominate or the radiation field is isotropic and Planckian. 
Such conditions prevail to a good approximation deep in stellar atmospheres, 
but the strict validity of LTE can not be assumed for the observable layers.
Instead, statistical equilibrium is established.
The non-linear interdependency of radiation field and level populations
is the essence of the non-LTE problem that in principle needs to be solved.

Nevertheless, stellar atmospheres are described well by the
condition of LTE for many practical applications.
In such a case non-LTE line formation assuming a LTE model atmosphere is
sufficient to produce the synthetic spectra required for the comparison with
observation.

The calculation of reliable occupation numbers for the energy levels 
involved in the transitions to be investigated is a prerequisite for an
accurate analysis of spectra. Good agreement between model spectra and
observation can be achieved when the
following conditions are simultaneously fulfilled in the non-LTE computations:
{\sc 1})~the local temperatures and particle densities, i.e.~the
atmospheric structure, are accurately known,
{\sc 2})~the radiation field is realistic, 
{\sc 3}) all relevant radiative and collisional processes are taken into
account, and 
{\sc 4})~high-quality atomic data are available.

Items {\sc 1}) and {\sc 2}) require a realistic physical model of the
stellar atmosphere and thus an {\it accurate atmospheric parameter
determination}. This will be the central topic here.
Items {\sc 3}) and {\sc 4}) are related to the model atoms for the non-LTE 
calculations, hence a careful construction of comprehensive and robust model
atoms is also required (see Przybilla, this volume).
Shortcomings in any of the four conditions result in
increased uncertainties of the analysis.

\subsection{Models and Codes}
Here we briefly recall the models and codes that provide the
basis for the examples discussed in Sect.~\ref{examples}.
We use a hybrid approach for the non-LTE line-formation computations.  
These are based on line-blanketed plane-parallel, homogeneous and hydrostatic 
LTE model atmospheres calculated with {\sc Atlas9} (Kurucz~\cite{kur93b}). 
Line blanketing is realised via consideration of Opacity Distribution Functions (ODFs,
Kurucz~\cite{kur93a}), where the too high iron abundance originally used by 
Kurucz is compensated by ODFs with an appropriately reduced metallicity.
This allows to approach modern values for the solar abundance (e.g.~Asplund et
al.~\cite{asplund05}), as iron contributes $\sim$50\% of the line opacity. 
Alternatively, {\sc Atlas12} (Kurucz~\cite{kurucz96}) is used for models 
with abundance peculiarities, allowing for opacity sampling.
The model atmospheres are held fixed in the following non-LTE calculations.

Non-LTE level populations and synthetic spectra are computed with recent versions of
{\sc Detail} and {\sc Surface} (Gid\-dings~\cite{gid81}; Butler \&
Giddings~\cite{but_gid85}; both updated by K. Butler). {\sc Detail} solves the 
coupled radiative transfer and statistical equilibrium equations employing the 
Accelerated Lambda Iteration scheme of Rybicki \& Hummer~(\cite{rh91}). 
This allows even complex ions to be treated in a realistic way. Synthetic
spectra based on the resulting non-LTE populations are computed with 
{\sc Surface}, using refined line-broadening data. 

Detailed tests have shown that this hybrid non-LTE approach is  
consistent with full non-LTE calculations, but faster, for our cases of interest,
BA-type supergiants (PBBK06) and OB-type
dwarfs and giants (NP07). It also allows comprehensive 
model atoms to be employed. This
can provide crucial improvements over full
non-LTE calculations, which are necessarily based on simplified model
atoms. An overview of model atoms adopted 
for our purposes is given in PBBK06 and PNB08.  

\section{Requirements on Observed Spectra}\label{observations}
The advent of high-resolution spectrographs with large wavelength 
coverage, large telescopes and automated data reduction pipelines
facilitates relatively easy access to high-quality spectra. 
However, some issues remain that 
introduce systematic errors into the quantitative analysis.

Continuum normalisation is one of the important issues, typically
not achieved in a satisfactory way for echelle spectra with available 
data reduction pipelines alone. Cross-checks with lower-resolution long-slit
spectra are recommended. A customised data reduction for order merging and
rectification (see e.g.~Hensberge~\cite{hensberge07}) becomes often
necessary when broad spectral features, like the hydrogen Balmer lines, 
are to be analysed. Uncertainties in the normalisation can 
lead to erroneous parameters from Balmer line analysis, such as for effective
temperature for the later-type stars or surface gravity for early-type stars.

The modelling and analysis techniques should be tested on slowly-rotating
stars of similar spectral type as the objects of interest.
High-resolution and high-S/N spectra of such `standard stars' are required 
to identify line blends. These may go unnoticed at lower spectral resolution 
or at higher rotational velocities, affecting analyses with
incomplete linelists. Spectra of rapidly rotating stars are
challenging, since rotational smearing can merge closely-spaced lines into a
`pseudo-continuum' below the true continuum. Line strengths may therefore be
systematically underestimated. Full spectrum synthesis can facilitate an
analysis, if continuum windows are found. 
The signal-to-noise (S/N) ratio can become one of the main sources of 
systematics, making continuum-definition difficult and weak lines inaccessible.
Severe limitations to the quality of line fits may be imposed~by~low~S/N.  
Some contribution of the observational material to the total uncertainty
budget for stellar parameters and elemental abundances can be
expected. However, well-selected targets and careful data acquisition and reduction can
help to avoid unnecessary systematics (see Sect.~\ref{errors} for more examples).

\begin{figure}[!t]
\vspace{-5mm}
\hspace{1.7cm}
\includegraphics[width=.82\linewidth]{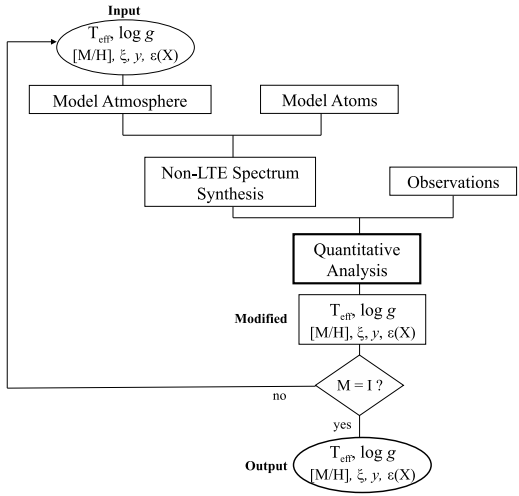}
\vspace{-10mm}
\caption{Schematic diagram of a self-consistent spectral
analysis methodology. Note the iterative nature of the procedure. Numerous sources 
of systematic error need to be eliminated in every aspect of the process to
achieve high accuracy. Spectral indicators for the quantitative 
analysis depend on the type of star.}
\label{diagram}
\end{figure}

\section{Self-Consistent Spectral Analysis}\label{analysis}
Here we discuss our recommended spectrum analysis technique. 
Well-known concepts are assembled into a procedure that allows multiple
stellar parameter indicators to be brought into agreement simultaneously. In combination
with a determination of element abundances this facilitates the
observation to be reproduced with a high degree of realism. The method works best
when all lines of the Balmer series, the helium lines and many
transitions of the metals are utilised. Spectral coverage at near-IR
wavelengths is an asset. The method is organised as an iterative
procedure, as sketched in Fig.~\ref{diagram}.

\begin{figure}[!t]
\begin{center}
\includegraphics[width=.989\linewidth]{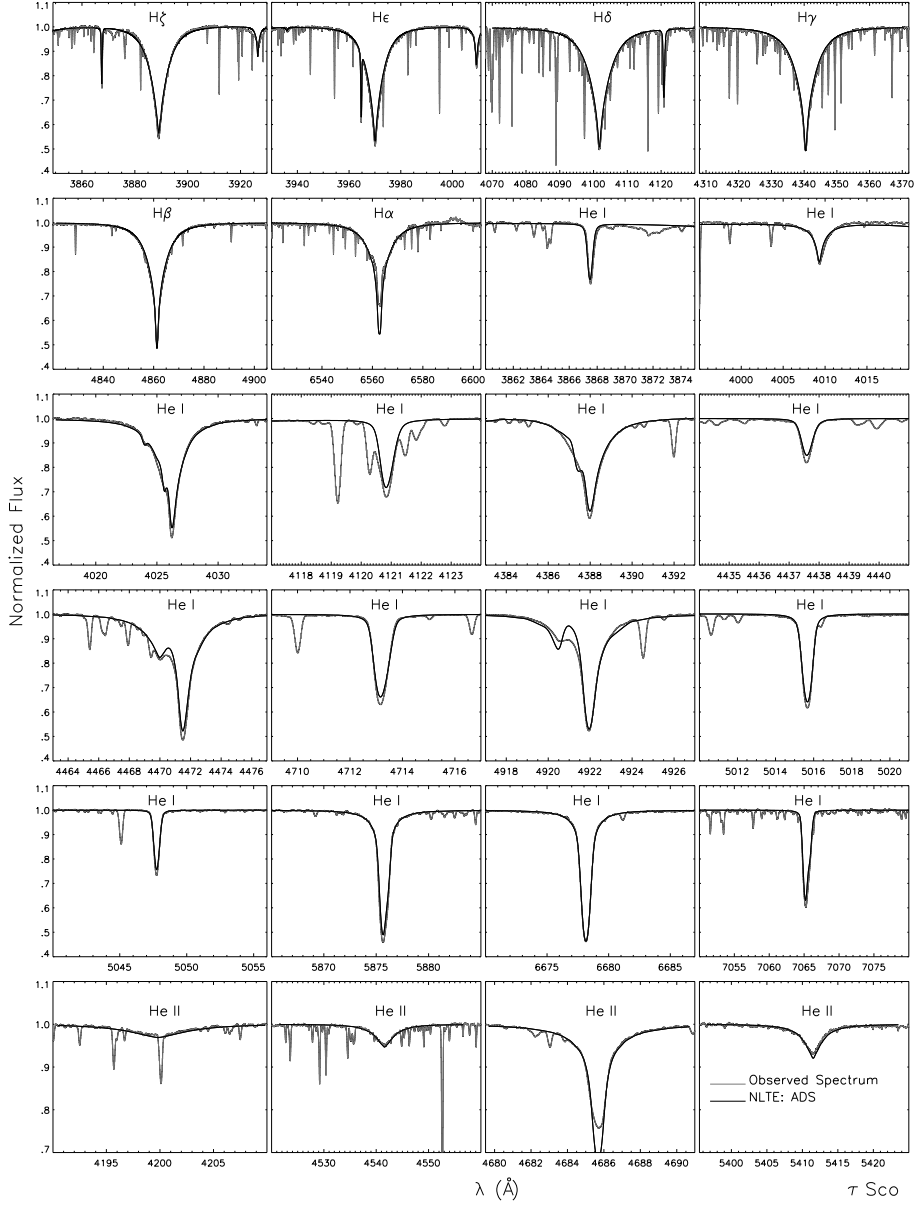}
\end{center}
\vspace{-8mm}
\caption{Simultaneous fits to most observable hydrogen and helium lines in
the optical spectrum of $\tau$\,Sco (B0.2\,V). Note that the He\,{\sc i/ii} 
ionization balance is established. The only exception is the
strong He\,{\sc ii}\,$\lambda$4686\,\AA~line which is affected by a 
weak stellar wind that is not accounted for in our calculations. From NP07.}
\label{HHEOB}
\end{figure}

\begin{figure}[!t]
\begin{center}
\includegraphics[width=.987\linewidth]{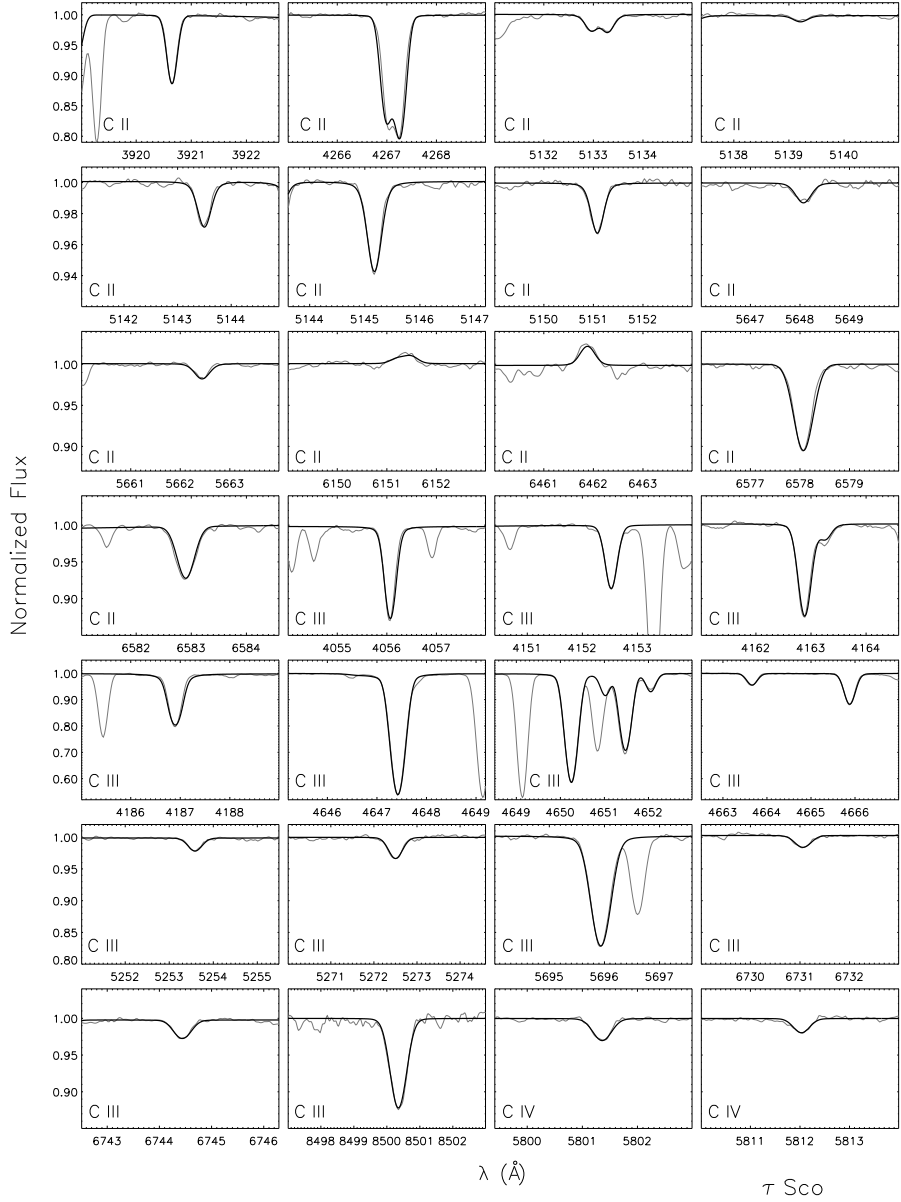}
\end{center}
\vspace{-8mm}
\caption{Simultaneous fits to most observable carbon lines in 
the optical spectrum of $\tau$\,Sco (B0.2\,V).
Every line fit adopts a slightly different abundance value, 
resulting in a 1$\sigma$ statistical uncertainty of 0.12\,dex from the
line-to-line scatter.
The C\,{\sc ii/iii/iv} ionization equilibrium is reproduced within 
these constraints. From NP08.
}
\label{COB}
\end{figure}

\begin{figure}[!t]
\begin{center}
\includegraphics[width=0.95\linewidth]{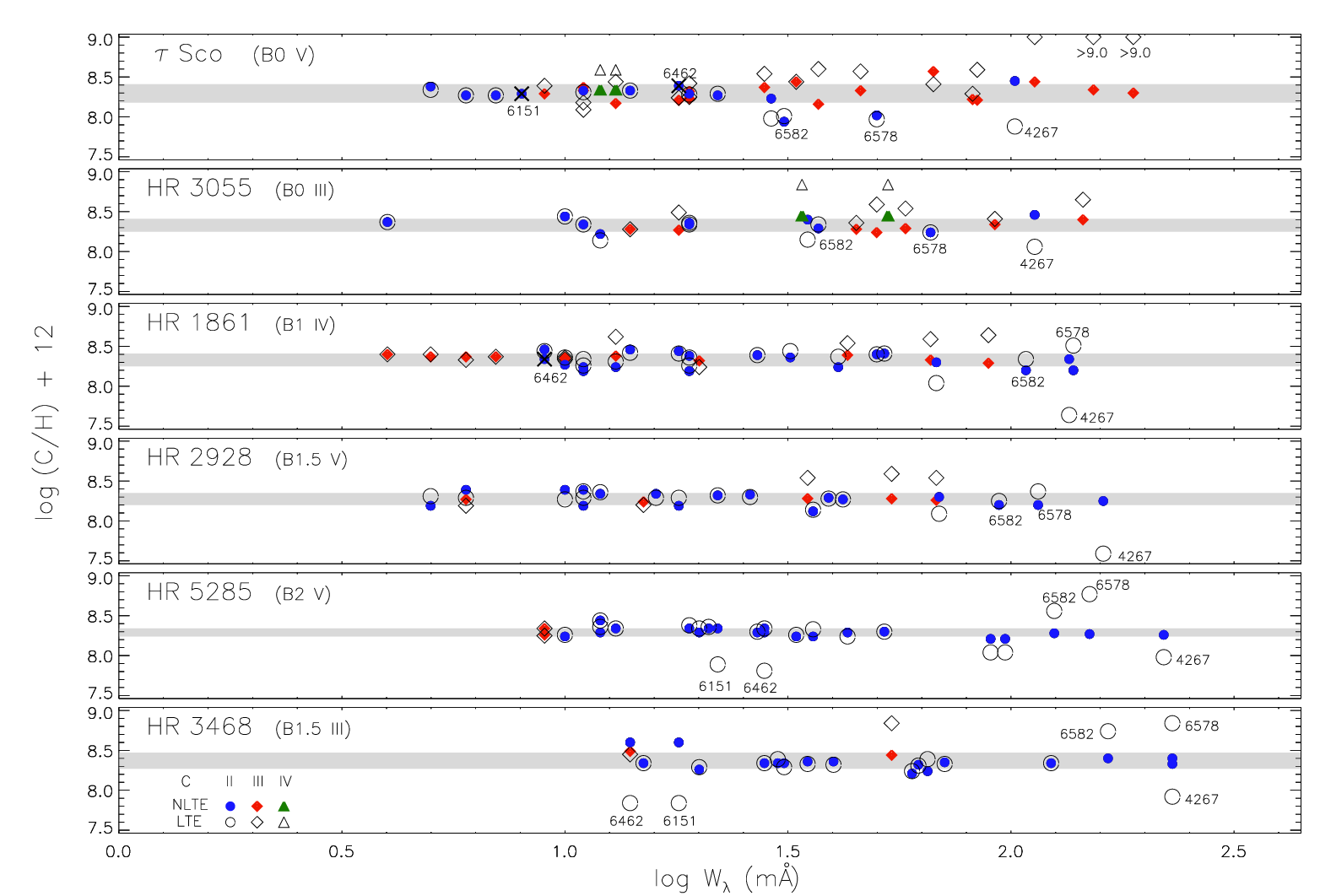}
\end{center}
\vspace{-6mm}
\caption{Non-LTE (filled symbols) and LTE (open symbols) carbon abundances
derived from line profile fits to individual lines of six early B-type stars 
in the solar neighbourhood. Single (circles), double (diamonds) and 
triple-ionized (triangles) carbon lines are analysed. Note that non-LTE abundances
show a lower scatter than the LTE results. The carbon ionization equilibrium
is reproduced consistently in non-LTE. From NP08.}
\label{abund}
\end{figure}

Good initial estimates for effective temperature, $T_\mathrm{eff}$,
surface gravity, $\log g$, He abundance, $y$, microturbulent velocity, $\xi$,
and metallicity, [M/H] (elemental abundances, $\varepsilon$(X)) are crucial 
to avoid many iteration cycles. Standard methods like photometric calibrations 
(see e.g.~Smalley~\cite{smalley05}) plus estimates for the
remaining quantities can be employed to gain initial approximations to start 
individualised calculations. Also, `secondary' parameters like
$\xi$ or $y$ have to be treated consistently in all steps of the
computations to avoid the introduction of systematic errors.   
The comparison of the resulting synthetic spectrum with the observed spectrum gives
indications as to how the input parameters should be
changed to improve the next iteration. Alternatively, and more
efficiently, pre-computed model grids may be utilised to perform line fits with 
a least-square algorithm (e.g.~{\sc Fitprof}, Napiwotzki~\cite{napi99}). 
Parameter estimates are obtained from a $\chi^2$-minimisation. The first aim 
is to reproduce the 
observed hydrogen and helium  spectra (see e.g.~Fig.~\ref{HHEOB}), the two elements 
constituting the bulk of the stellar material. Usually, reasonable 
values for $T_\mathrm{eff}$, $\log g$ and $y$ may be obtained for early-type 
stars. An advantage for the practical work is that the Stark-broadened hydrogen and 
helium lines are not highly sensitive to microturbulence and metallicity, except
close to the Eddington limit (see Sect.~\ref{errors}). 

\begin{figure}[!t]
\begin{center}
\includegraphics[width=0.88\linewidth]{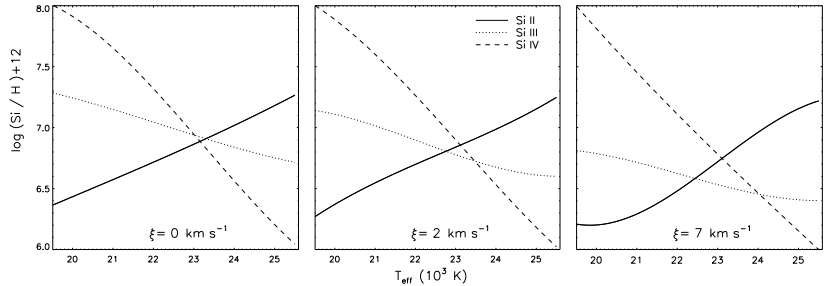}
\end{center}
\vspace{-6mm}
\caption{Effective temperature determination for the LMC star NGC\,2004-D15
via the silicon ionization equilibrium.
The Si\,{\sc iii} lines are strong, hence the
derived silicon abundance depends on microturbulent velocity, $\xi$, while
lines of the other two ions are less-sensitive to the choice of $\xi$
because of their weakness. Different
$T_\mathrm{eff}$ is derived from Si\,{\sc ii/iii} and Si\,{\sc iii/iv}
for ill-chosen $\xi$, e.g. an overestimation of $\xi$ by 7\,km\,s$^{-1}$ gives $\Delta
T_\mathrm{eff}$\,$\approx$\,2000\,K. The Si\,{\sc ii/iii/iv} ionization
equilibrium is established only for the correct $\xi$. From Nieva~(2007).}
\label{si-ie}
\end{figure}

When a good fit to the H and He lines is achieved -- $T_\mathrm{eff}$ and
$\log g$ should be accurate to better than about 5\% and 0.1--0.2\,dex,
respectively -- the procedure commences to 
consider metal lines. Ionization equilibria, i.e.~the requirement that lines
from different ions of an element have to indicate the same elemental
abundance, facilitate a fine-tuning of the previously derived parameters. 
Elements that show lines of three ionization stages in the 
spectrum are most valuable as they provide simultaneous constraints on both 
$T_\mathrm{eff}$ and $\log g$. Examples are C\,{\sc ii/iii/iv} 
(Figs.~\ref{COB} and \ref{abund}) or Si\,{\sc ii/iii/iv} in early B-type stars
(Fig.~\ref{si-ie}). Typically, however, lines from only two ionization stages
of an element are present in the spectrum, allowing only to constrain combinations of
$T_\mathrm{eff}$ and $\log g$. Another indicator is then required for the parameter 
determination, e.g.~the previously analysed H/He lines, or a second ionization equilibrium. 
We recommend the use of multiple ionization equilibria for the
parameter determination as the redundancy of information helps to
minimise~systematics. Iron ionization equilibria are in particular useful as
they put good constraints also on metallicity, [M/H]\,$\simeq$\,[Fe/H]. 

Simultaneously, the microturbulent velocity needs to be 
constrained to values more accurate than the initial estimates at this stage. 
This is done by plotting abundances from the line analysis
versus equivalent width $W_{\lambda}$ (see
Fig.~\ref{abund} for final results).
A slope in the results indicates that the microturbulent velocity has to be
adjusted. But there is much more to gain from such a plot:
an offset of abundances from two ions indicates that the
$T_\mathrm{eff}$/$\log g$ combination requires some fine-tuning. An
excessive scatter in abundances points to remaining inconsistencies,
e.g.~with basic model assumptions (LTE in Fig.~\ref{abund}) or with the 
model atom. 

The importance of a simultaneous determination of microturbulent velocity with 
other parameters -- which is facilitated
only iteratively -- is shown in Fig.~\ref{si-ie}. This visualises the dependency 
of the resulting $T_\mathrm{eff}$ to the adopted 
value of $\xi$ when only one ionization equilibrium of silicon is considered, 
i.e.~Si\,{\sc ii/iii} or Si\,{\sc iii/iv}. Different values are indicated
for an ill-chosen microturbulent velocity. The problem can be solved via use of multiple 
ionization equilibria, e.g.~by bringing Si\,{\sc ii/iii/iv} simultaneously
into agreement, and by considering also other elements as explained above.
The goal is to derive the same value of $T_\mathrm{eff}$, $\log\,g$ and $\xi$ from 
all hydrogen, helium and the metal lines ($y$ may need to be adjusted only slightly at
this stage).  

Abundance analyses for elements not considered so far 
finalise the metallicity derivation. Usually, this brings no surprises for
the stellar parameter determination. Exceptions can be chemically-peculiar
stars, where our procedure is also useful (Przybilla et al.~\cite{PNT08}) 
but more iteration steps can be required.

The final set of parameters has high precision and high accuracy. Absolute
values of $T_\mathrm{eff}$ and $\log\,g$ show residual uncertainties as low
as $\sim$1\% and 0.05--0.10\,dex, respectively, and for absolute elemental abundances
an rms scatter of $\sim$10--20\%. Remaining major sources of systematic error 
can essentially be excluded after bringing so many observational constraints
into agreement. A self-consistent treatment of `secondary' parameters such as  
$y$, [M/H] and $\xi$ throughout the entire process, from the 
atmospheric structure to the line-profile calculation is decisive for
achieving this.
Non-LTE ionization equilibria are sensitive to even small details of the
calculations, so turning this sensitivity into a high-precision tool
requires avoidance of any inconsistencies. We will resume the discussion on
systematics in Sect.~\ref{errors}.

A limitation for the practical work is imposed by the iterative approach, the
method is time-consuming. It is not feasible at the present time to analyse large 
samples of spectra efficiently. There are several ideas as to how to speed up the 
process, e.g.~by analysis of a restricted but well-selected sample of
spectral lines, which has to cover a variety of weak to strong lines 
of multiplets from the different spin systems of an ion. The full line
sample is then analysed only in the
final step. Extensive use of grids of non-LTE model spectra for metals 
with automatised line-fitting algorithms will play a crucial r\^ole in
speeding-up the method, but require considerable computational resources to sample 
the multi-dimensional parameter space ($T_\mathrm{eff}$, $\log\,g$, $\xi$, $y$,
[M/H] and abundances) properly. Nevertheless, we are confident that the analysis 
methodology will be applicable to high-precision studies of larger samples of stars in
the near future.

\section{Examples of Quantitative Analyses}\label{examples}
O and B-type main-sequence and giant stars, and their evolved progeny, the B and A-type supergiants, 
constitute the objects with the simplest photospheric physics among the massive 
stars. They are unaffected by strong stellar winds like the hotter and 
more luminous stars or by convection and chromospheres like the cool stars. 
In order to use their full potential as tracers of stellar and
galactochemical evolution (Przybilla~\cite{P08})
improvements in the quantitative analysis had to be implemented with respect
to previous work. Comprehensive and robust model atoms for
non-LTE line-formation calculations were constructed. In parallel, the
self-consistent analysis technique described in Sect.~\ref{analysis} was
implemented. Here we discuss some practical aspects from the use of the
recommended analysis technique, giving special emphasis on the impact of 
non-LTE effects on both the analysis procedure and the results.
An adaptation of the technique for the analysis of other types of stars could be done by
analogy.

\begin{figure}[!t]
\begin{center}
\includegraphics[width=0.55\linewidth,angle=90]{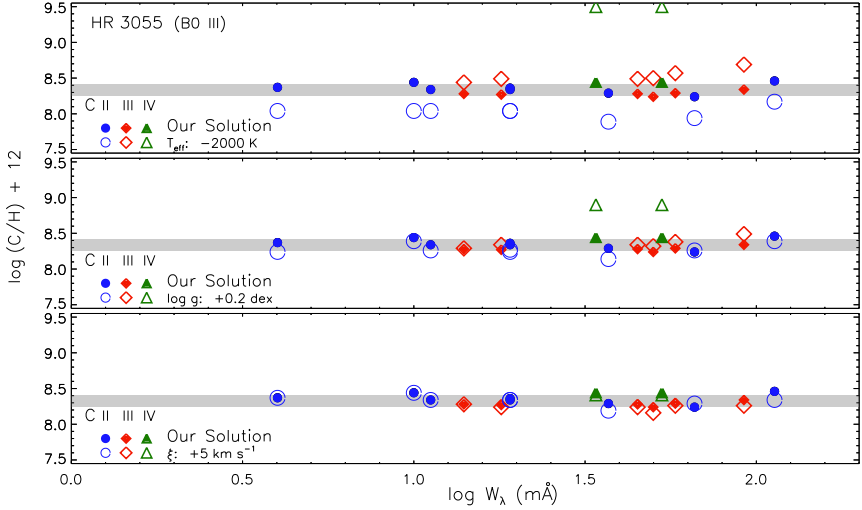}
\end{center}
\vspace{-6mm}
\caption{Example of systematic errors in carbon abundances 
caused by atmospheric parameter variations in a B0\,III star.
The offsets in parameters (values are indicated in the lower left corner of
each panel) are averaged discrepancies for studies using standard analysis techniques.
Incorrect parameters prevent consistent C\,{\sc ii/iii/iv}
abundances to be derived. 
Parameters determined from the carbon ionization equilibrium
(NP08) were recently confirmed by other metal ionization equilibria (PNB08).
>From NP08.
}
\label{param}
\end{figure}
\begin{table}[!ht]
\caption{Systematic errors in carbon
abundances (in dex) from individual lines 
caused by atmospheric parameter variations and the assumption of
LTE for line-formation calculations in the B0\,III star HR\,3055
 ($T_\mathrm{eff}$\,$=$\,31\,200$\pm$300\,K; $\log g$\,$=$\,3.95$\pm$0.05; 
$\xi$\,$=$\,8$\pm$2\,km\,s$^{-1}$).}
\label{param_abund}
 $$
\footnotesize{
 \begin{array}{lcccccccc}
 \noalign{}
  \hline
 \mathrm{Ion} & \lambda & W_{\lambda} &
 \varepsilon(\mathrm{C})_\mathrm{NLTE} & \Delta T_\mathrm{eff} & \Delta\log\,g & \Delta\,\xi  & \mathrm{LTE} \\
  &  \mathrm{(\AA)}  & \mathrm{(m{\AA})}  &  & -2000\,\mathrm{K} & +0.2\,\mathrm{dex}& +5\,\mathrm{km\,s}^{-1}& \\
 \hline\\[-2mm]
 \mathrm{C}$\,{\sc ii}$ &4267.2 & 113 & 8.46 &-0.33 &-0.11 &-0.16  &-0.40  \\[-.5mm]
           &5133.3 & 19 & 8.34 &-0.30 &-0.10 &~~0.00 &~~0.00  \\[-.5mm]
           &5143.4 & 10 & 8.44 &-0.40 &-0.05 &~~0.00 &~~0.00  \\[-.5mm]
           &5145.2 & 19 & 8.36 &-0.32 &-0.09 &-0.02  &~~0.00  \\[-.5mm]
           &5151.1 & 11 & 8.34 &-0.30 &-0.08 &~~0.00 &~~0.00  \\[-.5mm]
           &5662.5 & ~4 & 8.37 &-0.33 &-0.13 &~~0.00 &~~0.00  \\[-.5mm]
           &6578.0 & 66 & 8.24 &-0.40 &-0.15 &-0.10  &~~0.00  \\[-.5mm]
           &6582.9 & 37 & 8.29 &-0.30 &+0.02 &+0.05  &+0.05  \\[1mm]
 \mathrm{C}$\,{\sc iii}$ &4056.1& 45 & 8.28 &+0.21 &+0.06 &-0.04  &+0.08   \\[-.5mm]
             &4162.9 & 58 & 8.29&+0.28 &+0.09 &-0.03  &+0.25   \\[-.5mm]
             &4186.9 & 92 & 8.34&+0.35 &+0.15 &-0.08  &+0.07   \\[-.5mm]
             &4663.5 & 18 & 8.27 &+0.22 &+0.07 &-0.03  &+0.22   \\[-.5mm]
             &4665.9 & 50 & 8.24 &+0.26 &+0.08 &-0.08  &+0.35   \\[-.5mm]
             &5272.5 & 14 & 8.28 &+0.16 &+0.01 &~~0.00 &~~0.00   \\[1mm]
 \mathrm{C}$\,{\sc iv}$&5801.3 & 53 & 8.45& +1.06 &+0.46& -0.03  &+0.39\\[-.5mm]
          &5811.9 & 34 & 8.45 & +1.06 &+0.46& -0.03  &+0.39   \\[.5mm]
 \hline
 \end{array}
}
 $$
 \end{table}

Let us summarise the spectroscopic indicators for the parameter determination:
\begin{itemize}
\item $T_\mathrm{eff}$: all available hydrogen and helium lines \& ionization equilibria
\vspace{-3mm}
\item $\log g$: wings of all available hydrogen lines \& ionization equilibria
\vspace{-3mm}
\item $y$: all available helium lines
\vspace{-3mm}
\item $\varepsilon\mathrm{(X)}$: a comprehensive set of metal lines
\vspace{-3mm}
\item $\xi$: metals with numerous lines of different strength, slope 0 in $\varepsilon\mathrm{(X)}$ vs. W$_{\lambda}$
\vspace{-3mm}
\item $v\sin i$ and $\zeta$: from metal line-profile fits
\vspace{-3mm}
\item $\mathrm{[M/H]}$: abundance analysis
\end{itemize}
We see that the analysis requires high-resolution spectra with a near-complete wavelength 
coverage in the optical and reliable continuum rectification, preferentially at
high S/N. Additional constraints, when available, are recommended to be checked for
consistency, like spectral energy distributions (SEDs) or highly sensitive 
spectral lines in the near-IR (Nieva et al.~\cite{nieva09}; Przybilla, this volume).

\subsection{OB-type Main Sequence and Giant Stars}\label{OB}
Spectral indicators for $T_\mathrm{eff}$ and $\log g$ in late-O and early-B 
main sequence and giant stars are the hydrogen and helium lines. In addition,
several of the following ionization equilibria are used, depending on the 
temperature range: He\,{\sc i/ii}, C\,{\sc ii/iii/iv}, O\,{\sc i/ii}, Ne\,{\sc i/ii}, 
Si\,{\sc ii/iii/iv} and Fe\,{\sc ii/iii}.
Simultaneous fits to most observable H, He and C lines in a FEROS spectrum
are shown in Figs.~\ref{HHEOB} and \ref{COB}. In a hot star like $\tau$\,Sco
practically all the H and He lines show marked non-LTE strengthening.
Line cores as well as the wings are affected, see NP07 for details.
The He\,{\sc i/ii} and C\,{\sc ii/iii/iv} ionization balance is established 
simultaneously in non-LTE. This is impossible to achieve assuming LTE because of the
different sensitivity of the lines to non-LTE effects.
In the case of hydrogen and helium a unique model is employed for all fits. 
For carbon, because of the higher sensitivity of the lines to the chemical 
abundance, every line fit employs a slightly different abundance value. The
resulting non-LTE abundances for this element show a low rms scatter, and even
more so for other stars (see Fig.~\ref{abund}).

The effects of variations in $T_\mathrm{eff}$, $\log\,g$ and
$\xi$ on carbon abundances from individual lines in a B0\,III star are shown
in Fig.~\ref{param}. They have been quantified for
representative lines in Table~\ref{param_abund}. The offsets in parameters are 
averaged discrepancies for studies using standard analysis techniques like
photometric indicators. An offset in a parameter prevents consistent 
C\,{\sc ii/iii/iv} ionization equilibrium to be achieved, even when the model atom 
is highly reliable. Displacements of the average abundances for the different
ions occur in such a situation. This can introduce unnoticed 
systematics to abundance results whenever only one ion is studied for
poorly-constrained stellar parameters.
The effects are largest for $T_{\rm eff}$ variations and lowest for
modifications of $\xi$, in particular for the weaker lines.
Non-LTE effects can be the dominant factor for the abundance determination
for some lines (e.g.~C\,{\sc ii}\,$\lambda$4267\,{\AA} or C\,{\sc
iii}\,$\lambda$4666\,{\AA}, see
Table~\ref{param_abund}), but in many cases a reliable parameter
determination is more important than accounting for deviations from LTE.
Both non-LTE weakening and strengthening can occur for different lines of 
the same ion, and non-LTE effects may dominate even weak lines, 
contrary to common expectation. The best examples are the weak 
C\,{\sc ii}\,$\lambda\lambda$6151 and 6462\,{\AA} emission lines
(Fig.~\ref{COB}), which remain unexplained by LTE line formation. 
\begin{figure}[!t]
\begin{center}
\includegraphics[width=0.48\linewidth]{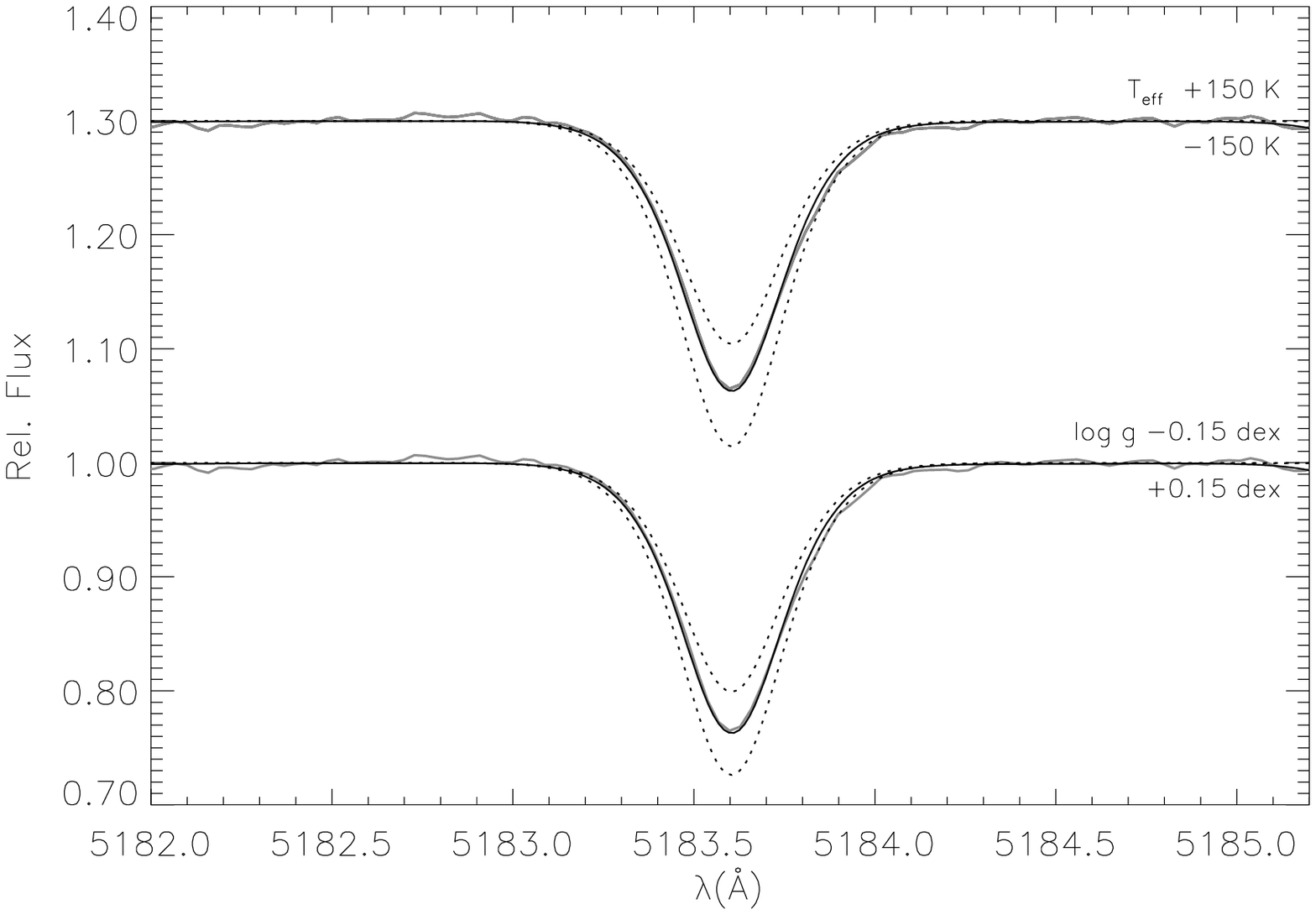}
\includegraphics[width=0.48\linewidth]{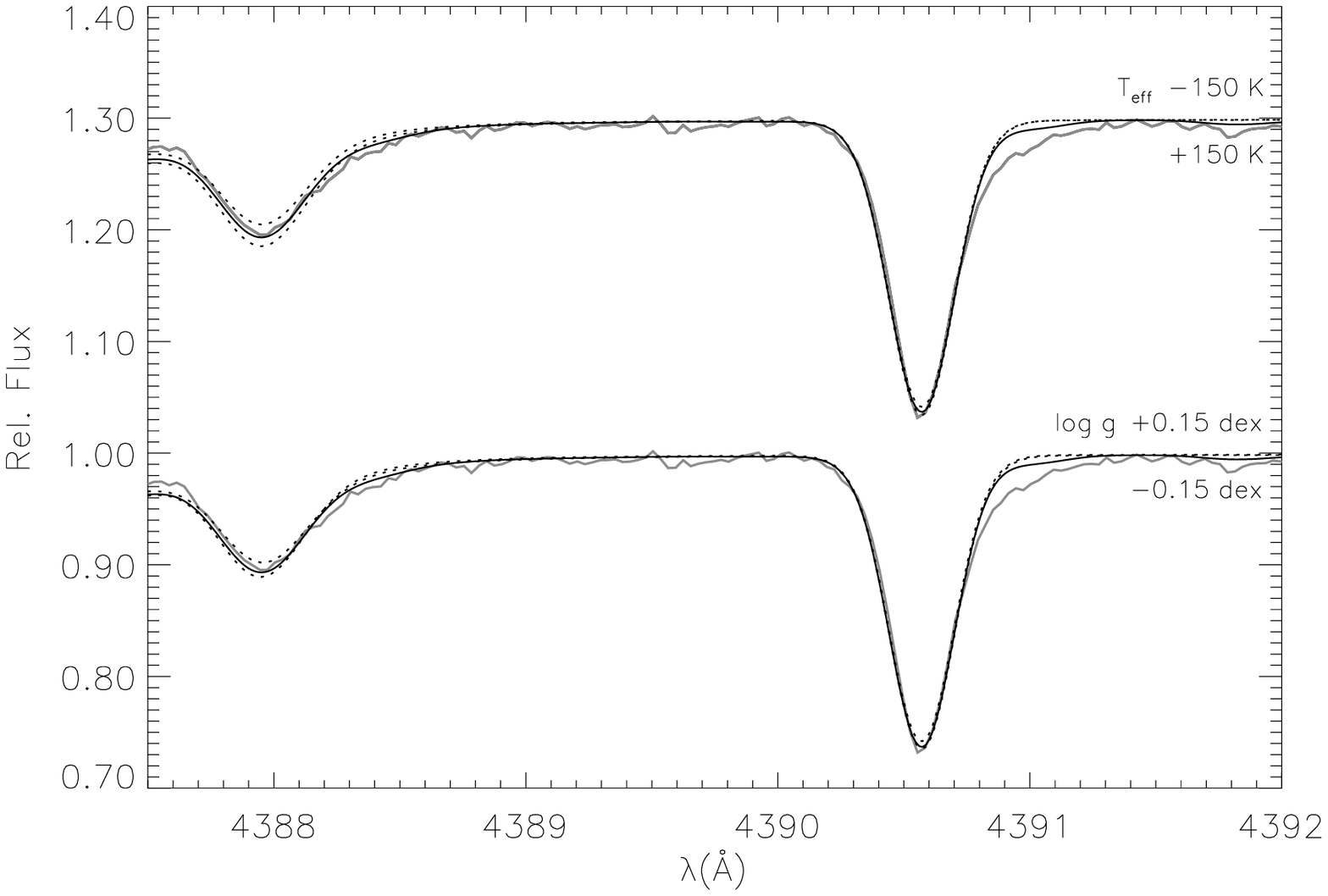}
\includegraphics[width=0.48\linewidth]{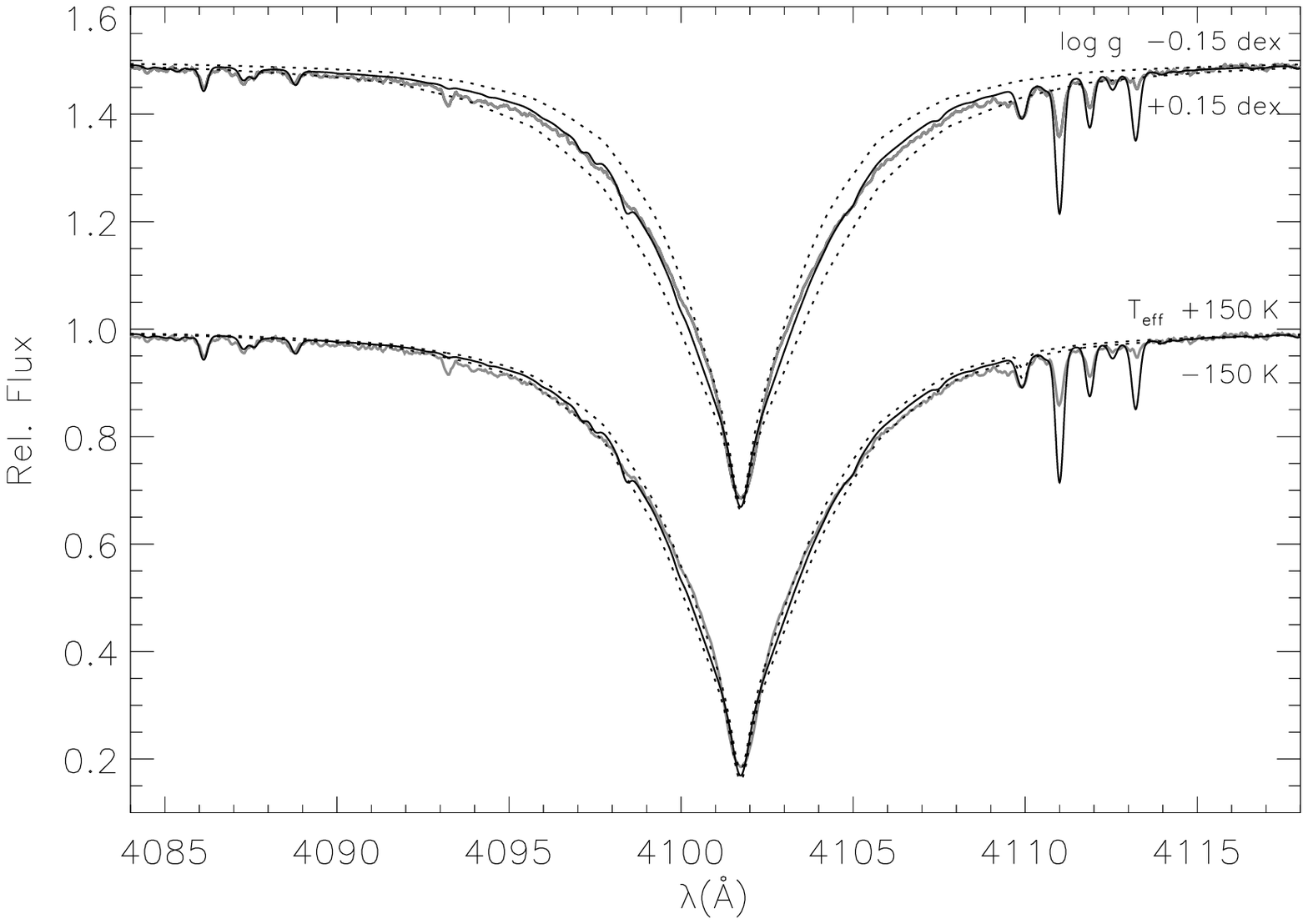}
\hspace{2.5mm}
\includegraphics[width=0.48\linewidth]{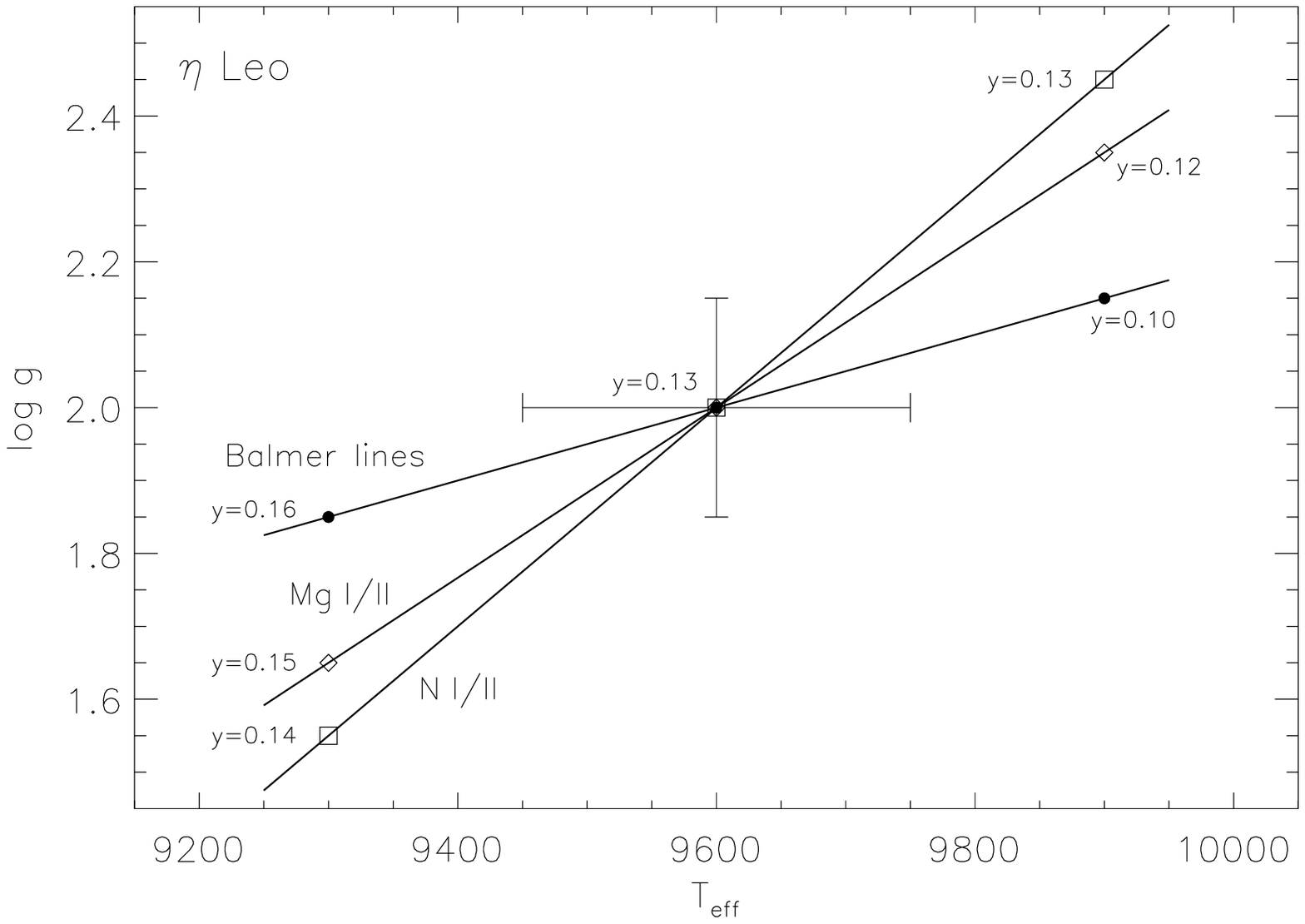}
\end{center}
\vspace{-8mm}
\caption{$T_\mathrm{eff}$ and $\log g$ effects on magnesium and hydrogen
lines, and the final solution of the parameter determination 
accounting also for the nitrogen ionization equilibrium
and helium abundances in $\eta$\,Leo, (A0\,Ib). 
Upper left panel: Mg\,{\sc i} lines are 
sensitive to $T_\mathrm{eff}$ and $\log g$ variations. 
Upper right panel:  Mg\,{\sc ii} lines are not
sensitive to these parameters. Lower left panel: sensitivity of
H$\delta$ to $T_\mathrm{eff}$ and $\log g$ variations. 
Lower right panel: final solution as the intersection of the different loci
for the H lines and Mg and N ionization equilibria with parametrised He
abundance in the $T_\mathrm{eff}$--$\log g$ plane. From PBBK06.}
\label{TEFFLOGGBA}
\end{figure}
\begin{figure}[!t]
\begin{center}
\includegraphics[width=0.70\linewidth]{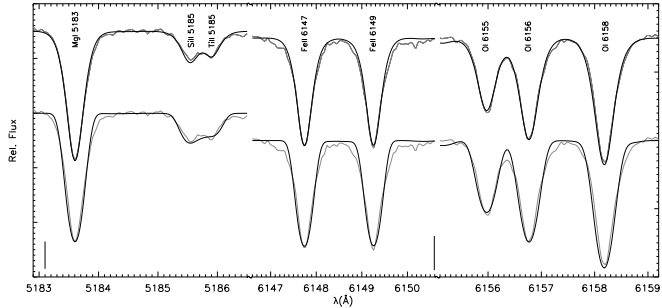}
\end{center}
\vspace{-8mm}
\caption{Determination of macroturbulent ($\zeta$) and projected equatorial rotation ($v\sin i$) velocities
in $\eta$\,Leo. 
The upper comparison between observation (thick) and spectrum synthesis 
(thin line, for finally derived mean abundances) takes into account $\zeta$
and $v\sin i$, while the lower comparison assumes a pure rotational
profile. From PBBK06.}
\label{macBA}
\end{figure}
\subsection{BA-type Supergiants}\label{examplesBA}
Spectral indicators for $T_\mathrm{eff}$ and $\log g$ in late-B and early-A 
supergiants are the Stark-broadened hydrogen lines and several non-LTE ionization 
equilibria. We utilise at least two of the following: C\,{\sc i/ii}, N\,{\sc i/ii}, 
O\,{\sc i/ii}, Mg\,{\sc i/ii}, Si\,{\sc ii/iii} or S\,{\sc ii/iii}. Examples for 
the (high) sensitivity of the Mg\,{\sc i/ii} lines and a Balmer line in an A0\,Ib 
supergiant to $T_\mathrm{eff}$/$\log g$ variations are shown in
Fig.~\ref{TEFFLOGGBA}. Each indicator allows to construct a curve in the
$T_\mathrm{eff}$--$\log g$ plane for which the observed features are
reproduced well. Note that the helium abundance has to be varied 
accordingly in this process, i.e. the observed He\,{\sc i} lines need to be
reproduced simultaneously, as the atmospheric structure of such a luminous object 
reacts noticeably to changes of the mean molecular weight of the plasma. The intersection
of all loci determines the final parameters (see Fig.~\ref{TEFFLOGGBA}).

\begin{figure}[!t]
\begin{center}
\includegraphics[width=0.95\linewidth]{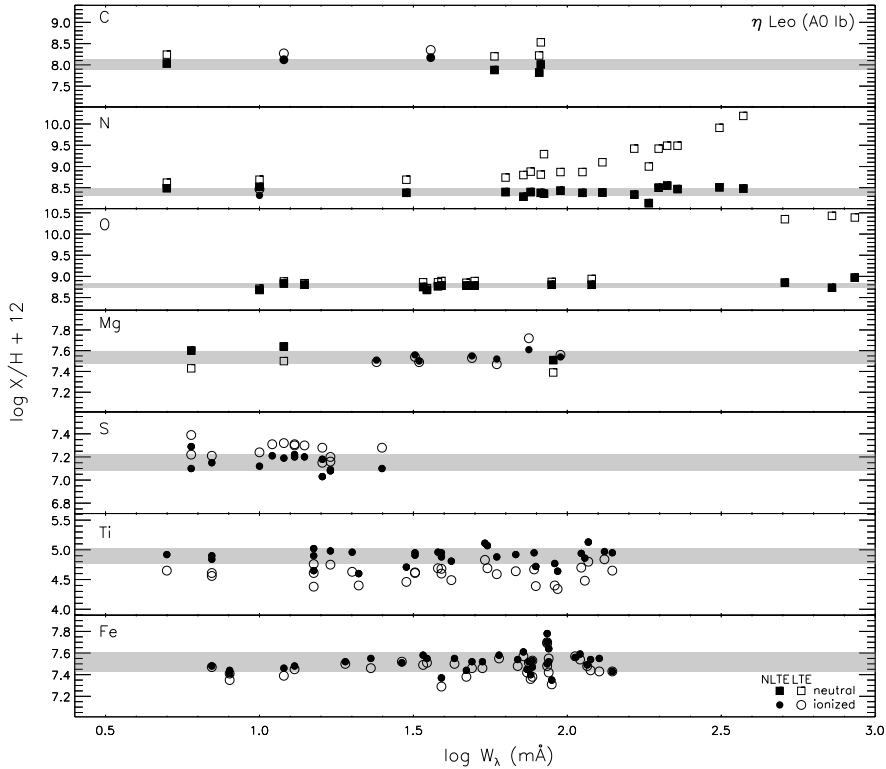}
\end{center}
\vspace{-8mm}
\caption{Elemental abundances from individual spectral lines
as a function of equivalent width, see the legend for symbol encoding.
%non-LTE (full symbols) and LTE (open
%symbols) for neutral (boxes) and single-ionized species (circles).
The grey bands cover the 1$\sigma$-uncertainty around the non-LTE averages.
Proper non-LTE calculations reduce the line-to-line scatter and remove
systematic trends. The same value for microturbulent velocity is adopted
in all cases. From~PBBK06.}
\label{microBA}
\end{figure}

Detailed line-profile fits are much more powerful than analyses of
equivalent widths (i.e. of only an integrated quantity). For line fitting,
additional parameters need to be known, the projected equatorial rotational velocity,
$v \sin i$, and the macroturbulent velocity, $\zeta$, which we determine within our
spectrum synthesis approach. The rotational and macroturbulent
broadening profiles are convolved with the synthetic spectrum, and $v \sin i$ 
and $\zeta$ are varied until the observed profiles are matched. This allows high
accuracy of these parameters to be obtained. Both, regions with unblended 
or with blended lines are useful for the derivation (see Fig.~\ref{macBA}).
Alternatively, Gray's Fourier transform technique may be employed (see 
e.g.~Sim\'on-D\'iaz \& Herrero~(\cite{SiHe07}) for applications to early-type stars).

Figure~\ref{microBA} is analogous to Fig.~\ref{abund}, but for different
elements in an A0\,Ib star. Besides indicating a proper choice of
microturbulence, which is found to be the same for all elements 
from the non-LTE analysis,
from slope zero of $\varepsilon$(X) vs. $W_{\lambda}$, also a match of the 
ionization equilibria (C\,{\sc i/ii}, N\,{\sc i/ii} and Mg\,{\sc i/ii}) is
found. Moreover, the comparison with LTE abundances shows that non-LTE
effects are ubiquitous, i.e.~they have to be accounted for in every step of
the analysis. The non-LTE analysis finds homogeneous abundances from all lines in the
element spectra, which show little scatter around the mean value (less than
in LTE). Systematic trends of abundance with line-strength found in LTE, 
e.g.~for N\,{\sc i} or O\,{\sc i}, are absent in non-LTE. In other cases, for
S\,{\sc ii}, Ti\,{\sc ii} and Fe\,{\sc ii}, the non-LTE abundances are
systematically shifted relative to LTE. Contrary to common assumption 
significant non-LTE abundance corrections are found even in the weak-line 
limit. These can exceed a factor $\sim$2 for lines with 
$W_{\lambda}$\,$<$\,10\,m{\AA}, in particular for more luminous 
objects with amplified non-LTE effects. At the other extreme, for the
strongest oxygen or nitrogen lines, abundance corrections can
amount to a factor $\sim$50, or even more. 

\begin{figure}[!t]
\begin{center}
\includegraphics[width=.99\linewidth]{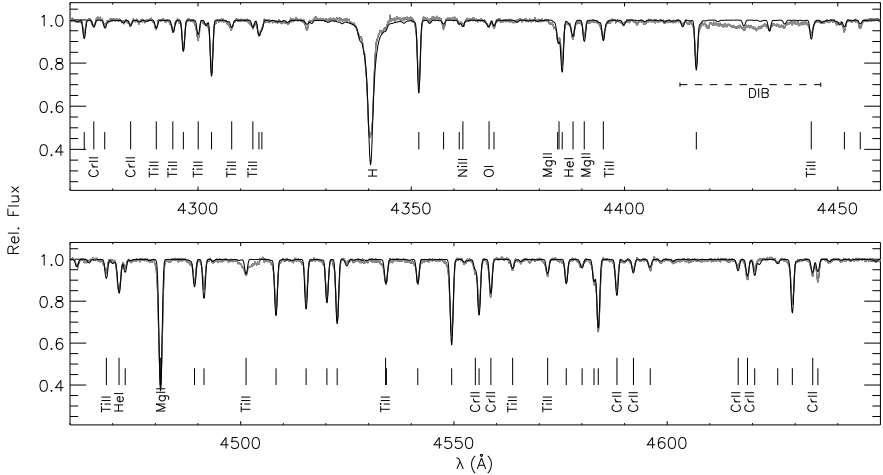}\\[-1cm]
\end{center}
\vspace{-8mm}
\caption{Global spectrum synthesis for HD\,92207 (A0\,Iae) resulting from the 
successful quan\-titative spectral analysis. From PBBK06.}
\label{SPECBA}
\end{figure}

The fruit of such a comprehensive stellar parameter and abundance analysis 
is a synthetic spectrum, which should reproduce the observed spectrum. 
A comparison of model with observation is the final test.
Very good agreement can indeed be obtained even for extreme objects, see
Fig.~\ref{SPECBA}. Note that a few discrepancies can remain, like those imposed by the
presence of diffuse interstellar bands (DIBs).

\section{Summary of Common Sources of Systematic Errors}\label{errors}
Every step in the quantitative spectral analysis is susceptible to systematic
errors. Here we briefly list the most common sources of
systematics that affect the final uncertainties of stellar parameters and elemental
abundances. We concentrate on systematics encountered in the study of
massive stars and, when possible, give recipes for how to prevent them. The
overview should be useful also for work on other types of stars, but can 
include additional complications, such as convection in
cool stars or magnetic fields, which are beyond the scope of the present
discussion.

\begin{figure}%[!ht]
\begin{center}
\includegraphics[width=0.49\linewidth]{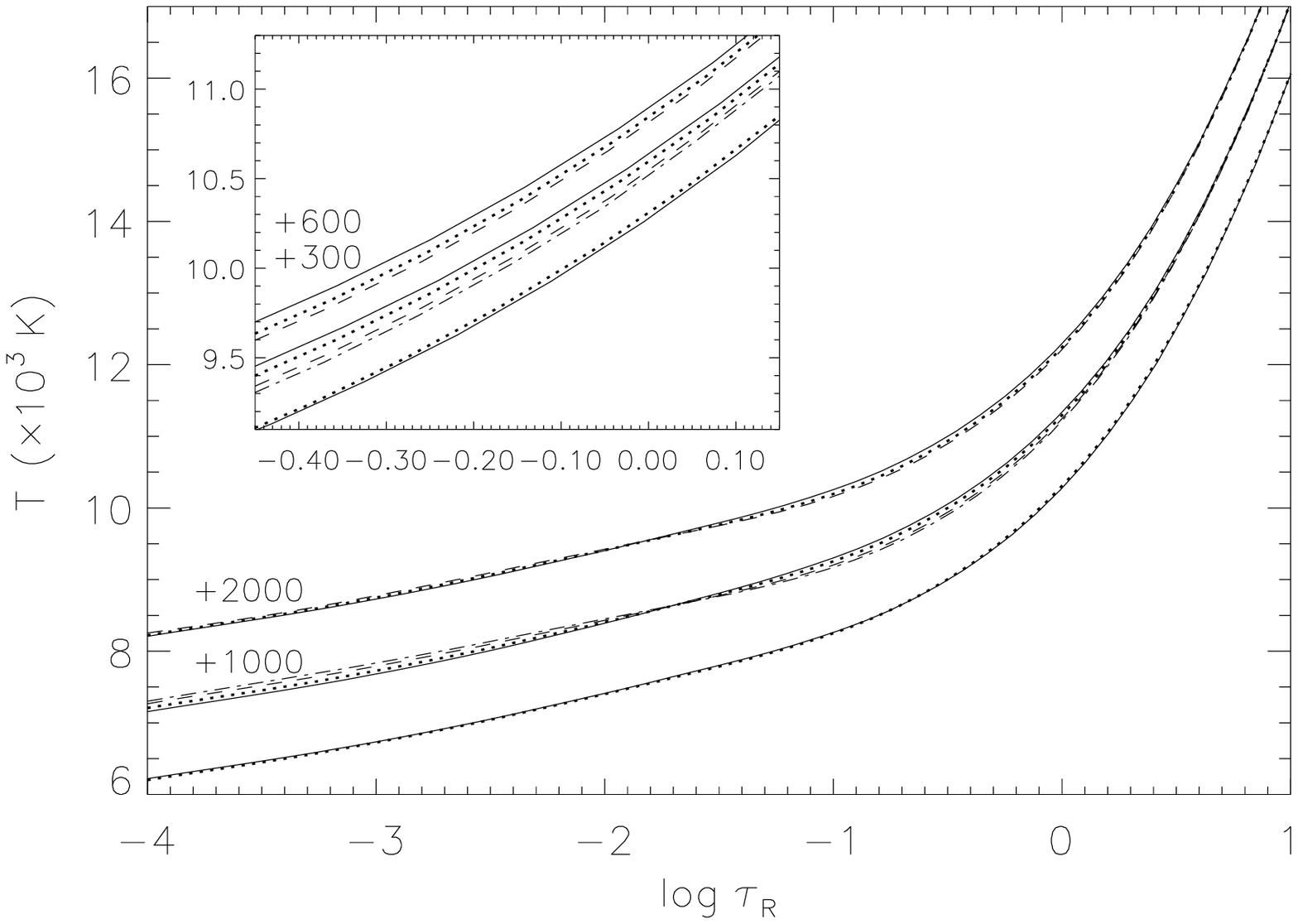}
\hfill
\includegraphics[width=0.49\linewidth]{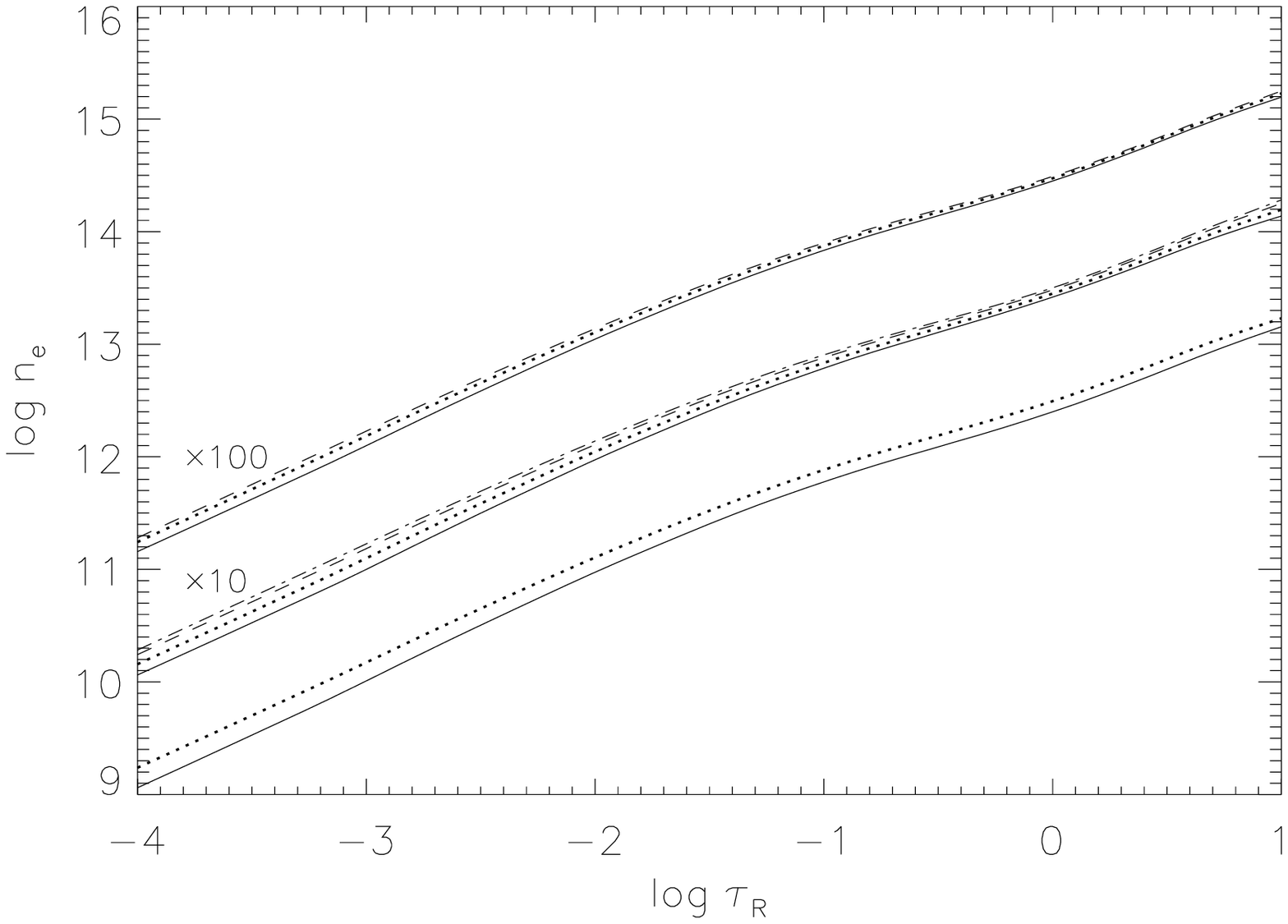}
\end{center}
\vspace{-6mm}
\caption{
Left panel: variations of the atmospheric temperature structure in response
to changes of helium abundance (lower), and effects of metallicity (middle) and
microturbulence (upper set of curves) on the atmospheric line blanketing.
The comparison is made for stellar parameters
matching those of an A0\,Iae star. Only one parameter is modified at each
time: $y$\,=\,0.089 (solar value, full line) and $y$\,=\,0.15 (dotted line);
use of ODFs with $[$M/H$]$\,=\,0.0, $-$0.3, $-$0.7 and $-1.0$\,dex (full,
dotted, dashed, dashed-dotted lines); use of ODFs with $\xi$\,=\,8, 4 and
2\,km\,s$^{-1}$ (full, dotted, dashed lines). For clarity, offsets by
$+$1000\,K and $+$2000\,K have been applied to the middle and upper set of curves,
respectively. In the inset, the formation 
region for weak lines is enlarged. Right panel: analogous to the left panel, but for the
atmospheric electron density. From PBBK06.}
\label{atmospheric}
\end{figure}

\begin{figure}%[!ht]
\begin{center}
\includegraphics[width=0.49\linewidth]{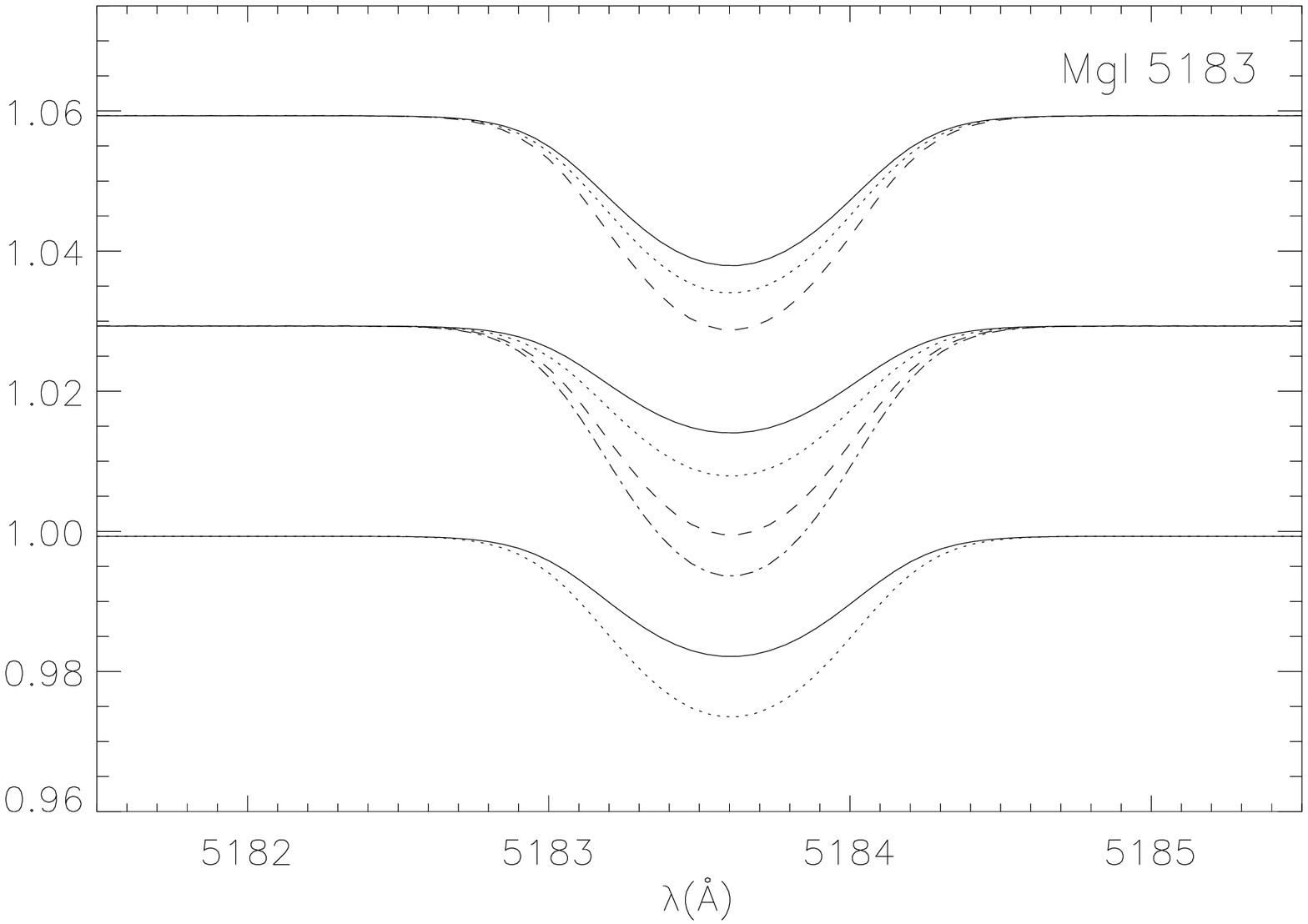}
\hfill
\includegraphics[width=0.49\linewidth]{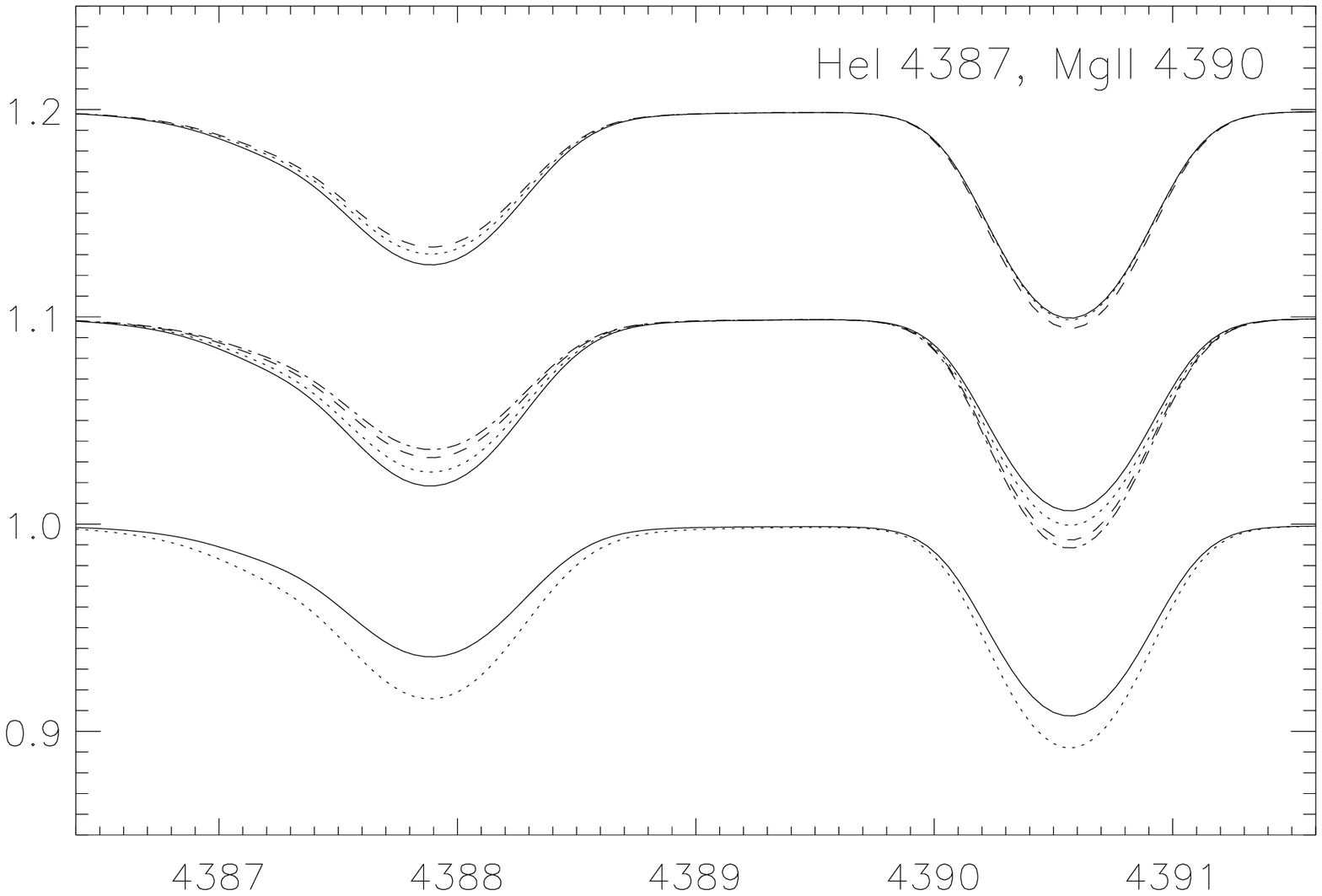}\\
\includegraphics[width=0.49\linewidth]{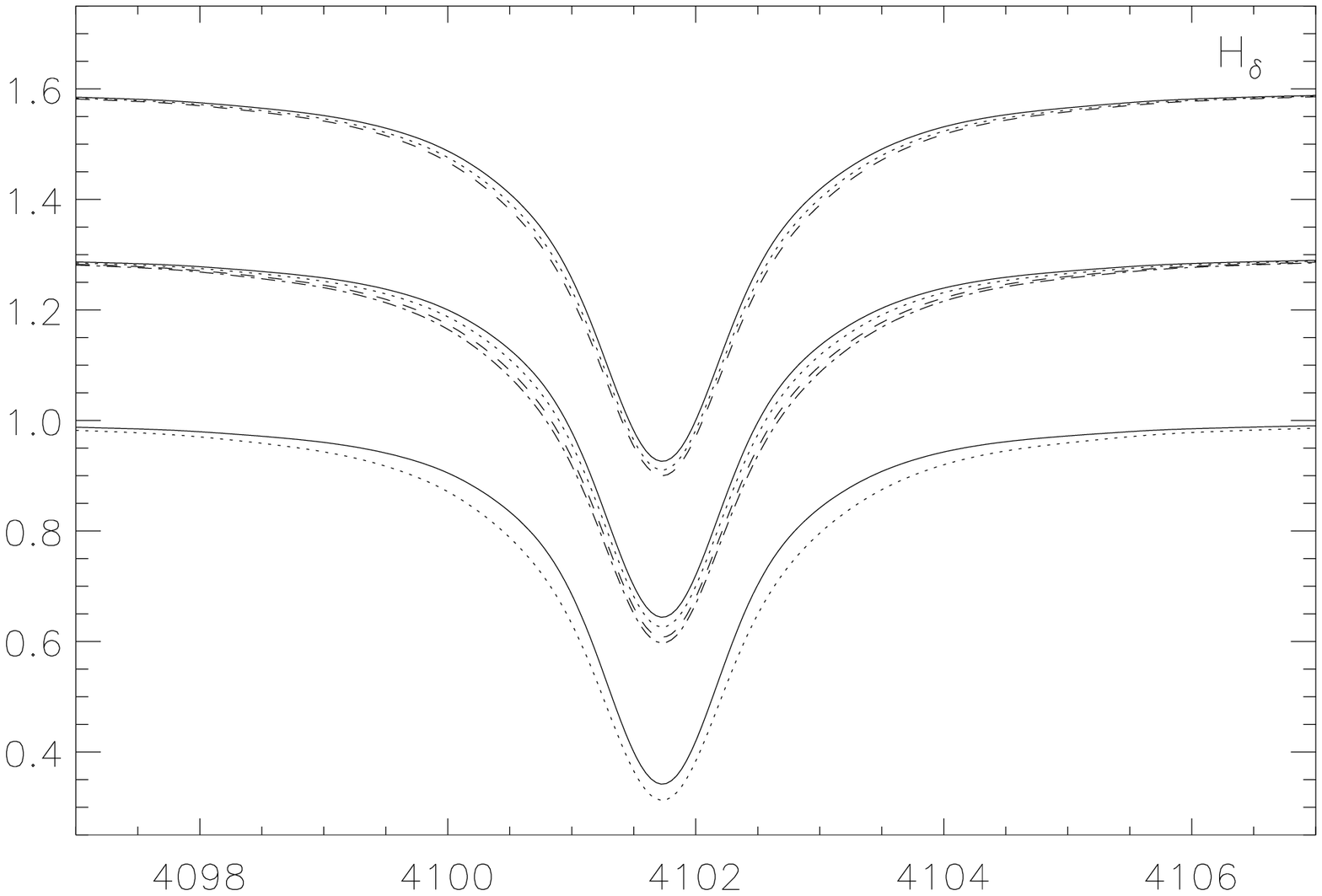}
\end{center}
\vspace{-6mm}
\caption{
Effects of varying atmospheric helium abundance (lower), metallicity (middle) and
microturbulence (upper set of curves) on the profiles of diagnostic lines in
an A0\,Iae supergiant, using the models discussed in Fig.~\ref{atmospheric}.
Note that the line profile changes are due to the effects of
modified mean molecular weight and line blanketing alone. Values of e.g. $T_{\rm eff}$,
$\log g$ and the magnesium abundance remain unchanged.
For clarity, offsets have been applied to the different sets of curves.
>From PBBK06.}
\label{profiles}
\end{figure}

\noindent{\bf Model atmospheres and line formation.} 
Atmospheric structures/SEDs computed with {\it full} non-LTE or {\it hybrid} non-LTE
methods are equivalent for studies of unevolved OB-type stars (NP07) and BA-type
supergiants (PBBK06). This is {\it not} a source of systematics for these
cases, but this needs to be tested for other types of objects. Crucial for
comparisons is that the same abundances are used in both approaches.
Abundance values need to be checked explicitly for model comparisons, 
e.g.~for `solar' abundances, as the solar `standard' has changed with time.
Secondary parameters like helium abundance, microturbulent velocity and metallicity
can affect atmospheric structures via their impact on the mean molecular
weight or on line-blanketing/blocking effects. Examples of the effects
to be expected on the structure of a loosely-bound atmosphere of a star 
close to the Eddington limit and on the resulting synthetic spectra are
shown in Figs.~\ref{atmospheric} and \ref{profiles}, respectively. In such a case the
effects are not negligible at all (see PBBK06 for a detailed discussion). These 
parameters need to be treated in a consistent way.\\[1mm]
{\bf Line blocking.} The presence of numerous lines in a spectrum affects the
radiation field. The radiation is blocked within the lines, and
redistributed into neighbouring continuum bands because of flux conservation.
Typically, the flux in the UV, where many lines are present, is
diminished and more radiation is emitted in the optical.
Non-LTE line-formation calculations need to account for this line blocking
effect, e.g. by using ODFs or opacity sampling techniques. The `softer'
radiation field yields less ionization, affecting all ionization equilibria.\\[1mm]
{\bf Model atoms.} Most model atoms for non-LTE calculations are
based on input atomic data ({\it ab-initio} data and approximations) as available 
in the early 1990s. It is worthwhile to check for improvements on the modelling 
whenever new data becomes available. As an example, different carbon model atoms 
can yield discrepancies in abundance analyses up to 0.8\,dex for
strategic lines (NP08).\\[1mm]
{\bf Line-broadening theory.} Progress made on line-broadening
theory has provided a wealth of detailed broadening parameters over the past
years. Detailed tabulations like those of Stehl\'e \& Hutcheon~(\cite{StHu99})
for the Stark-broadening of hydrogen line profiles should be favoured over
simpler approximations for a reliable stellar parameter determination. For
cool stars a detailed treatment of the broadening of spectral lines by 
neutral hydrogen collisions is highly important (e.g.~Barklem et
al.~\cite{barklem00a, barklem00b}).\\[1mm]
\noindent{\bf Composite spectra.}
Binary stars are common, and spectra of apparently single stars may turn out
to be of composite nature. Light from the fainter, secondary star may not be
accompanied by radial velocity variations for wide binaries, or for chance
alignments of stars when observing dense
fields. Standard analysis methods for single stars applied to a spectrum where the light 
from both stars is detected can give erroneous results throughout. Inspection of
hydrogen or helium lines for asymmetries, or for the presence of (weak) features not fitting
a star's spectral type may be helpful to identify such cases. In addition,
signatures like UV or IR excess should be checked in case of doubt.\\[1mm]
{\bf Quality of spectra.} Continuum normalisation and local continuum definition
are major sources of systematics for the analysis of spectra with
S/N\,$\le$\,50. The abundance determination in stars rotating
at intermediate velocities ($v\sin i\sim 50-150\,\mathrm{km\,s^{-1}}$) is
already limited to 0.2--0.3\,dex in accuracy at this S/N. Rapidly-rotating
stars ($v\sin i >
150\,\mathrm{km\,s^{-1}}$) or low-resolution spectra impose even more complications 
because metal line blends lower the actual continuum. Hence, the
abundances can be systematically underestimated and the accuracy is typically 
limited to 0.3--0.4\,dex (see Korn et al.~\cite{korn05}).\\[1mm]
{\bf Effective temperatures.} $T_\mathrm{eff}$ estimated from photometry can differ 
by more than 10\% (NP08) from that determined with a self-consistent spectroscopic method. 
Spectroscopic determinations via ionization equilibria
are a powerful technique only when the model atoms are reliable. 
In addition, {\it one} ionization equilibrium alone does not provide accurate
constraints because of dependencies on other variables 
(see  Fig.~\ref{si-ie} for the case of silicon) and consequences for carbon abundances
in Fig.~\ref{param} and Table~\ref{param_abund}.
Therefore, only the simultaneous use of multiple ionization equilibria
provides a precise and accurate $T_\mathrm{eff}$ determination.\\[1mm]
\noindent {\bf Surface gravity.} The common use of {\it only one}
Balmer line as $\log\,g$ indicator does not allow for consistency checks. 
Instead, all available H lines and in addition metal ionization equilibria should be
considered for an accurate determination. Moreover, the neglect of non-LTE effects 
on the Balmer lines, as often found in the literature,
can lead to systematic errors: increasing with
$T_\mathrm{eff}$ up to $\sim$0.2\,dex in $\log\,g$ around 35\,000\,K, see
e.g.~Table~\ref{param_abund} for effects on the carbon analysis.\\[1mm]
\begin{figure}%[!ht]
\begin{center}
\includegraphics[width=0.47\linewidth]{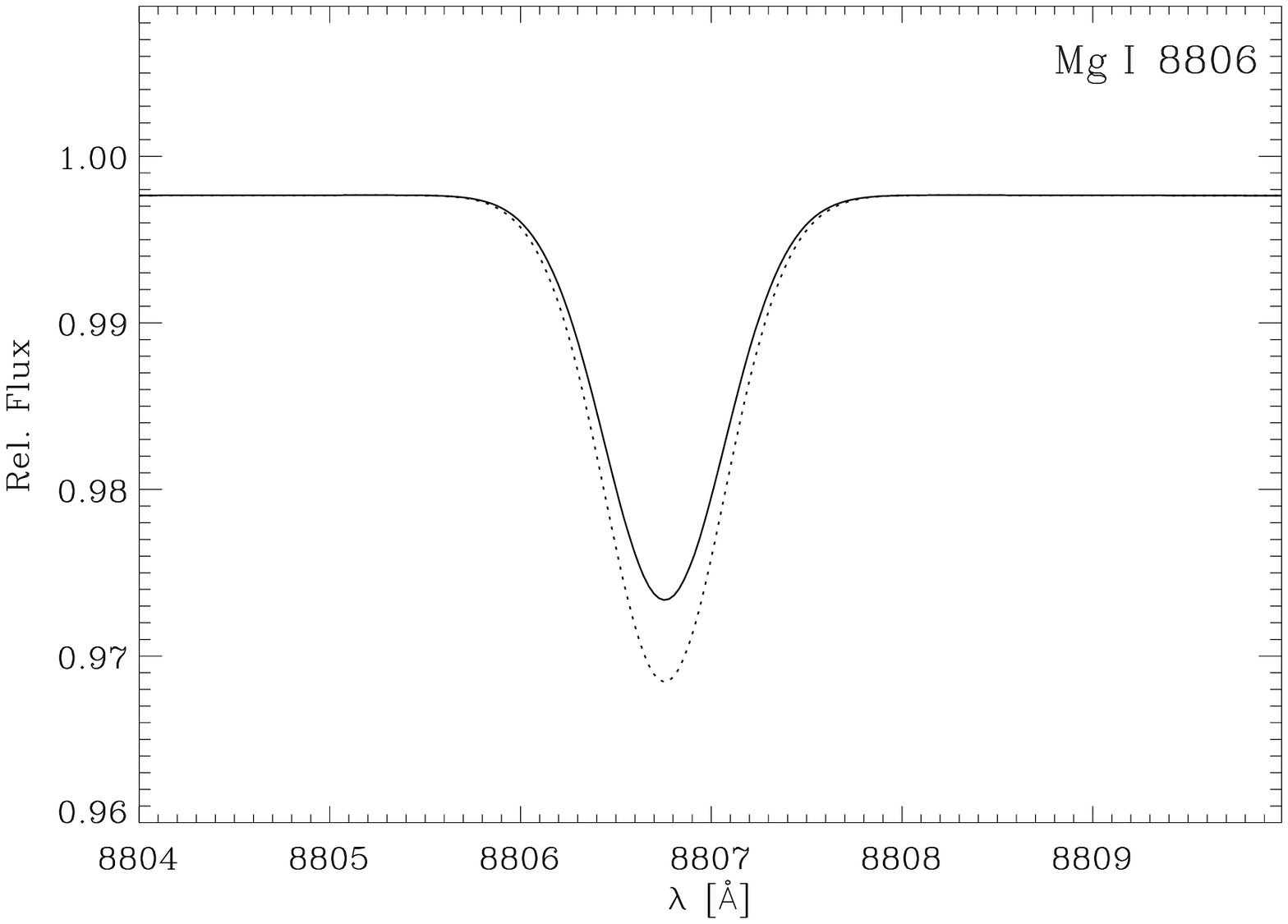}
\hfill
\includegraphics[width=0.47\linewidth]{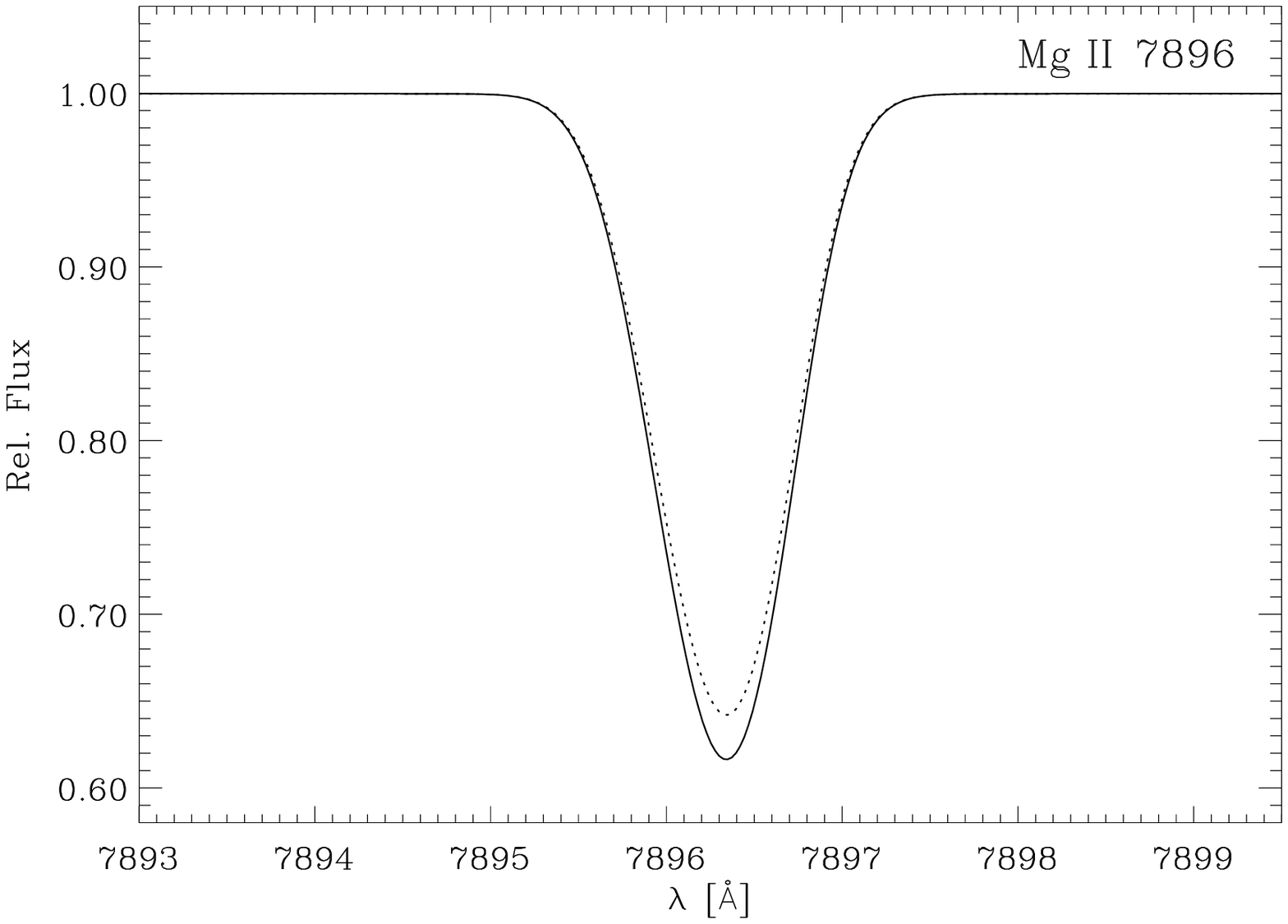}
\end{center}
\vspace{-6mm}
\caption{Theoretical line profiles for magnesium lines in $\eta$\,Leo,
assuming $\xi$\,$=$\,10\,km\,s$^{-1}$. The predictions
differ when microturbulence is accounted for self-consistently in the
statistical equilibrium and radiative transfer calculations (full line) and
when it is considered only in the formal solution (dotted), which is the
standard. From Przybilla et al.~(\cite{Przybillaetal01}).}
\label{microturbulence}
\end{figure}
{\bf Microturbulence.} This quantity is often derived from lines of only one
ion of one element (e.g.~O\,{\sc ii} in early-type stars) and in extreme 
cases from only one multiplet (e.g.~Si\,{\sc iii}\,$\lambda\lambda$4552--4574). 
Cross-checks with different species are mandatory when only a few lines are 
measurable (like in fast rotators) to avoid large uncertainties.
Identical values for $\xi$ are expected to be obtained in a proper non-LTE
analysis of different ions/chemical species, see Sect.~\ref{examplesBA}. 
Microturbulent velocities in excess of the sound speed are suspicious.
Often neglected is the effect of microturbulence on the radiative
bound-bound transition rates in the non-LTE computations. The strengths of
lines, their shape and their formation depth may be affected (see
Fig.~\ref{microturbulence} for an example).\\[1mm]
{\bf Spectral line selection.} Abundances may depend on the selection of lines used 
in the analysis when the model atoms are not comprehensive: some multiplets may 
indicate systematically different abundances than others. Which lines should be
trusted for the analysis? Comparison with other model atoms may help, some
transitions may be found not to deviate from LTE. However, we recommend to
investigate the reasons for such discrepancies, and implementation of
improvements for the model atoms. In principle, all possible observed lines of an
atom/ion should be analysed (see NP08 and PNB08).\\[1mm]
{\bf Non-LTE `corrections'.} Non-LTE effects cannot be easily predicted.
They affect different lines in different ways (see Fig.~\ref{microBA}). 
Non-LTE line strengthening or 
weakening can occur, or lines may turn out to be `in LTE'. Non-LTE effects are not 
restricted to the stronger lines alone. For different plasma conditions the non-LTE effects 
change, see Fig.~\ref{abund}. Hence, adding or subtracting fixed `non-LTE abundance 
corrections' to LTE results may increase the systematics.\\[2mm]
{\bf Macroturbulence.} Macroturbulence is not considered in many studies.
This is important for proper $v\sin i$ determinations, in particular for apparently
slowly-rotating objects. An example is shown in Fig.~\ref{macBA}, where a $v\sin i$
of zero is derived when macroturbulence is accounted for and $\sim$13\,km\,s$^{-1}$ 
if it is not.

\section{Conclusions}
Highly-precise and accurate stellar parameters {\it can} be spectroscopically 
determined, with limiting uncertainties as low as $\sim$1\% in
$T_\mathrm{eff}$, $\sim$0.05--0.10\,dex in $\log\,g$ and $\sim$10-20\% in
elemental abundances (rms scatter). A self-consistent spectral analysis 
methodology using non-LTE line formation was presented that allows this to be 
achieved when typical systematics are avoided. Of crucial importance is to simultaneously bring multiple
spectroscopic indicators into agreement, which requires an iterative approach.
The method is much more time-consuming than standard approaches for the stellar
parameter determination, but it is worth the effort whenever highly-accurate
observational constraints are required for astrophysical applications.

%In the past years we have made great efforts to reduce uncertainties
%in quantitative spectral analyses of BA-type supergiants and OB-type
% main sequence and giant stars. 
%A self-consistent analysis in non-LTE, i.e.~account for several spectroscopic
%indicators -- hydrogen and helium lines and multiple metal ionization equilibria
%-- resulted in drastically reduced systematic 
%effects in the atmospheric parameter and elemental abundance determination.
%A large number of potential systematic errors was identified when comparing our 
%new models and self-consistent method with standard techniques.
%Here we have summarised the model and observational data requirements as 
%starting points for a proper spectral analysis. We further described 
%the self-consistent quantitative spectral analysis method that drives us 
%to achieve high accuracy in the final parameters and summarised the most common
%sources of systematic errors that can be prevented. 
%The whole procedure is not an easy task, but is worth when high precision
%stellar parameters and chemical abundances unbiased by systematic errors are
%needed for further applications, e.g. as constraints for numerous contemporary astrophysical models.

%%-----------------------------
%%      your bibliography
%%-----------------------------

\end{document}